\documentclass{article}
\usepackage[T1]{fontenc} % add special characters (e.g., umlaute)
\usepackage[utf8]{inputenc} % set utf-8 as default input encoding
\usepackage{ismir,amsmath,cite,url}
\usepackage{graphicx}
\usepackage{color}

\usepackage{booktabs}
\usepackage{amssymb}
\usepackage{comment}
\usepackage{multirow}
\usepackage{placeins}
\usepackage{lineno}
\usepackage{balance}
\usepackage{flushend}
\usepackage{microtype}
\usepackage{xcolor}

\usepackage{url}            % simple URL typesetting
\usepackage{booktabs}       % professional-quality 

\title{``More than words'': Linking Music Preferences and Moral Values through Lyrics}
\multauthor
{Vjosa Preniqi$^1$ \hspace{1cm} Kyriaki Kalimeri$^2$ \hspace{1cm} Charalampos Saitis$^1$} {
 $^1$ Centre for Digital Music, Queen Mary University of London, London UK \\
$^2$ ISI Foundation, Turin, Italy\\
{\tt\small \{v.preniqi, c.saitis\}@qmul.ac.uk,
kyriaki.kalimeri@isi.it,
}
}

\def\authorname{V. Preniqi, K. Kalimeri, and C. Saitis}

\usepackage[bookmarks=false,pdfauthor={\authorname},pdfsubject={\papersubject},hidelinks]{hyperref}
\begin{document}

\maketitle
\begin{abstract}
% CHARIS: the abstract is now 163 words. If we can change it next week, we should probably add one introductory sentence that summarises motivation and/or novelty
This study explores the association between music preferences and moral values by applying text analysis techniques to lyrics. 
Harvesting data from a Facebook-hosted application, we align psychometric scores of 1,386 users to lyrics from the top 5 songs of their preferred music artists as emerged from Facebook Page Likes.
We extract a set of lyrical features related to each song's  overarching narrative, moral valence, sentiment, and emotion. 
A machine learning framework was designed to exploit regression approaches and evaluate the predictive power of lyrical features for inferring moral values.
Results suggest that lyrics from top songs of artists people like inform their morality. Virtues of hierarchy and tradition achieve higher prediction scores ($.20 \leq r \leq .30$) than values of empathy and equality ($.08 \leq r \leq .11$),
%, in line with previous literature. 
%The prediction was only improved on average by 1\% when user demographics were considered.
while basic demographic variables only account for a small part in the models' explainability. 
% with respect to  moral values.
This shows the importance of music listening behaviours, as assessed via lyrical preferences, alone in capturing moral values.
We discuss the technological and musicological implications and possible future improvements.
% Prediction results show that the values related to Binding were the most predictable by our data (.65 \textit{AUROC}). Our findings suggest that song lyrics are informative of moral values. Lyrics attributes even outperform demographics when predicting Binding values. Therefore, this association, along with the most predictive lyrics features that we provide for each moral foundation, can be beneficial for improving personalized content in music recommendation systems. Further, we discuss the technical and musicological implications and possible future improvements.
\end{abstract}
\section{Introduction}\label{sec:introduction}

The field of music recommender systems has a lot to gain from the fields of music psychology and sociology \cite{porcaro2021diversity,laplante2014improving}, where researchers have found converging evidence that people listen to music that reflects their personality needs \cite{rentfrow2003re,greenberg2016song,nave2018musical,devenport2019predicting,anderson2021just} and helps express their values \cite{gardikiotis2012rock,swami2013metalheads,manolios2019influence}. 
For example, extroverted people tend to choose more energetic and rhythmic tunes, while listeners holding values of understanding and tolerance prefer more sophisticated and complex music.
Indeed, operationalising knowledge of how personality traits relate to listener taste and preferences has already been shown to improve music recommendations \cite{hu2011enhancing,jin2020effects} and to make them more diverse \cite{lu2018diversity}.
Yet personality dispositions alone may not suffice to explain, and thus model, our music listening behaviours. 

Aiming to advance an integrative view of the music listener, which may benefit music recommender system scenarios, we set to explore the less attended relation between moral values and music preferences. If personal values are conceived as intrinsic motivational goals, moral values reflect traits learned under the influence of society, culture, and religion, amongst others, which bond people together into groups.
Considering music as an evolved tool of social affiliation and bonding \cite{loersch2013unraveling,savage2021music}, it is reasonable to speculate that people may like certain music styles and genres because they provide stimuli that match their morality-related needs.

We further hypothesise that moral values are expressed more clearly in a verbal rather than non-verbal manner and examine their influence on musical taste through lyrics. 
When people listen to sung music, their preferences are driven by not only the audio content but also the content of lyrics \cite{demetriou2018vocals}.
Lyrics convey rich, multifaceted messages about societal issues such as love, life and death, but also political or religious concepts, often independently from melodic and other audio information \cite{ali2006songs}.
% A recent study of the BTS boy band fans has revealed that the lyrical messages have often helped listeners find comfort and reassurance by poising escapism and distraction while externalizing suppressed thoughts \cite{lee2021a}.
% \VP{The lyrical messages have often helped listeners find comfort and reassurance by poising escapism and distraction while externalizing suppressed thoughts \cite{lee2021a}.}
Lyrical messages can support listeners' mental health \cite{lee2021a}.
Nonetheless, little is known about whether lyrical information manifests links between psychological traits and music preferences \cite{qiu2019personality}.
To what extent are moral values reflected in the lyrics of one's favourite songs? 
Do lyrics predict the moral traits of listeners?

To tackle these questions, we used data from the \textit{LikeYouth.org} project, a Facebook-hosted application developed specifically for research purposes as a surveying tool and was mainly deployed in Italy. 
Upon providing their informed consent, participants completed validated psychometric questionnaires for personality, moral traits and basic human values, basic demographic information such as age and gender, while agreed to share their Page Likes (see \cite{urbinati2020young} for a detailed description of the complete dataset).
For the purpose of this study we only analysed moral values scores and Likes on music artist Pages.
Combining these with information from the \textit{genius.com} music database, we obtained the lyrics from the five most popular songs per artist.
We performed both sentiment \cite{hutto2014vader} and emotion \cite{mohammad2013crowdsourcing} analysis on the obtained lyrics, assessed their moral narratives employing the MoralStrength lexicon \cite{araque2020moralstrength}, and examined themes and overarching narratives through topic modelling \cite{blei2003latent}.

% Our study aims to shed light on the moral narratives in music, associating people's intrinsic moral foundations and music preferences, while taking into consideration emotion, sentiment, but also basic demographic features. 
We built a series of regression models that infer moral traits from lyrical content, demographics, and Likes-based features (e.g., artist popularity). 
Our findings show that peoples' worldviews and moral values are indeed reflected in their music preferences as modelled through lyrics, in line with recent literature \cite{preniqi2021modelling,messick2020role}. Extracting topic and moral features from lyrics specifically increased model performance over sentiment, emotion, and demographic features.

% Moreover, moral values aside to provide more actionable information they seem to be more differentiating with respect to personality traits~\cite{kosinski2014manifestations}.

We contribute to the growing literature studying the interplay between music and psychology, with findings that clearly link the preferences of people to artists and songs that are in line with their moral values.
Personalised recommendations for streaming on-demand music can be greatly enriched by including notions of moral worldviews about their listeners instead of only shallow psychological attributes~\cite{nave2018musical,porcaro2021diversity,manolios2019influence}. 
Such knowledge can be directly implemented in psychologically aware music recommender systems, improving music streaming services and contributing to listener wellbeing \cite{mejova2019effect}. 
On a different key, the relationship between moral worldviews and music preference is crucial to inform  communication experts about their choice of the most appropriate music piece to accompany a social campaign.

%
% \section{Theoretical Background}
\section{Background}

We operationalise morality via the Moral Foundations Theory (MFT) \cite{graham2013moral}, which expresses the psychological basis of moral reasoning in terms of five innate foundations, namely Care/Harm, Fairness/Cheating, Loyalty/Betrayal, Authority/Subversion, and Purity/Degradation. These can further collapse into two superior foundations: \textit{Individualising} (Care and Fairness), indicative of a more liberal perspective, and \textit{Binding} (Purity, Authority and Loyalty), indicative of a more conservative outlook.
% The five MFT traits are usually assessed through the Moral Foundations Questionnaire (MFQ) \cite{graham2011mapping}.
% two superior moral foundations: of \textit{Individualising}, compounded by Fairness and Care, which asserts that the basic constructs of society are the individuals and hence focuses on their protection and fair treatment; and of \textit{Binding}, which summarises Purity, Authority and Loyalty, and is based on the respect of leadership and traditions.
% MFT has been shown to work across different cultures

% Overall, these studies show that moral values are complex and thus harder to predict compared to personality traits or demographics. Nevertheless, they provide a realistic dimension of the possibilities of inferring moral traits for delivering better targeted and more effective interventions.

% \subsection{Moral Foundations and Music Preferences}

Moral foundations are considered to be higher psychological constructs than the more commonly investigated personality traits \cite{mcadams2006new}. They have been associated with attitudes towards complex situations such as politics \cite{miles2015morality, haidt2007morality}, climate change \cite{wolsko2016red}, and vaccination \cite{amin2017association, kalimeri2019human}. 
% Prior studies have pointed to a link between moral regards and personality traits \cite{noser2015dark, lewis2011left}. A study showed that Individualising foundations,
% characterised by concerns for righteousness and protection of individuals, were significantly associated with Agreeableness and Openness; Binding morals, mirroring appreciation of in-group order and aspirations for a pure life, were associated with Extroversion, and Conscientiousness. Whereas, Neuroticism was associated with both, Binding and Individualising foundations \cite{lewis2011left}. 

However, moral values have attracted less attention from music scientists.
% In recent literature, there are indications that music-induced negative emotions can worsen moral judgement \cite{ansani2019you}.
% that study did not rely on a psychometrically validated theory like MFT. 
Using data from an ad-hoc online survey comprising, among other items, MFQ scores and preferences ratings on 13 music genres, Preniqi et al.~\cite{preniqi2021modelling} found that people with higher levels of Binding foundations (e.g., more authoritarian individuals) tend to listen to country and Christian music, the lyrics of which often foster notions of tradition \cite{rentfrow2003re}. Those with lower levels of Binding traits tend to prefer music genres such as punk and hip-hop, where lyrics are known to challenge traditional values, and the status quo \cite{gardikiotis2012rock}. Individualising foundations were overall harder to predict (cf.~\cite{kalimeri2019predicting}).
Furthermore, including demographic information (e.g., age, gender, political views, education) improved MFT predictions marginally, indicating the ability of music preferences alone to explain one's moral values.

% Our findings suggest that musical preferences are quite informative of deeper psychological attributes; still there is space for improvement. For instance, we noticed that the care, fairness, and loyalty foundations are harder to predict. To this end we aim to explore musical content analysis, for instance, incorporating linguistic cues, and the moral valence scores as proposed by Araque et al. \cite{araque2020moralstrength,araque2021language} on lyrics to further improve the performance. 

% In our work, we propose a moral-driven approach to tackle individual music preferences focusing on linguistic cues while giving an additional lead towards better user understanding with the potential on improving future recommendation systems.

% \subsection{Music Preferences and Lyrics}

In the computational social science field, recent work has demonstrated the predictability of MFT traits from a variety of digital data, including gameplay \cite{kim2013moral}, smartphone usage and web browsing \cite{kalimeri2019predicting}. 
Moral values can also be explained by verbal data, as they can be more clearly communicated through thoughts and opinions \cite{kalimeri2019human, lin2018acquiring, araque2020moralstrength}. Several dictionary-based approaches for predicting moral values expressed in texts such as tweets and other social media posts have been proposed, including the Moral Foundations Dictionary \cite{graham2009liberals,lin2018acquiring} and the MoralStrength lexicon \cite{araque2020moralstrength}. Here we employ the latter to uncover moral narratives in song lyrics, which we then use to predict the moral traits of listeners.

% Research in linking musical taste with personality traits or personal and moral values has traditionally relied on genre preferences, obtained either through surveys (e.g., \cite{gardikiotis2012rock,preniqi2021modelling}) or, more recently, digital behaviour data from social media \cite{nave2018musical} and streaming services \cite{anderson2021just}.   
The relation between lyrics and music preferences has only recently started to receive attention across music and social psychology disciplines.
Some studies have suggested associations between the personality or mental health of songwriters and their lyrics \cite{greenberg2020self,lightman2007using}. 
On the listener side, neurotic individuals tend to listen to songs with more complex and less repetitive lyrics that express negative emotions \cite{shriram2021much,qiu2019personality}.
More conscientious individuals tend to prefer lyrics talking about achievements \cite{qiu2019personality} but also about love \cite{sust2022personality}.
Importantly, preferences for lyrics are found to be predictive of personality traits distinctly from audio or melodic preferences \cite{qiu2019personality,sust2022personality}.

% Applying lyrical content analysis to smartphone music listening data, 

% Sust and colleagues \cite{sust2022personality} applied lyrical content analysis to smartphone music listening logs

% Focusing on metal music sub-genres and using MFT alongside BIG5, 

Concerning moral values, in recent work, they have been found to explain a unique and significant portion of the variance in the lyrical preferences of different metal music sub-genre fans that was not already accounted for by personality traits \cite{messick2020role}. For example, preferring lyrics about celebrating metal culture and unity was related to higher levels of the Loyalty foundation and higher levels of extroversion.
In U.S.~popular music, an increase in lyrics related to self-focus and -promotion since the 1980s has been shown to manifest the increasing individualism of American society \cite{dewall2011tuning,mcauslan2018billboard}.

% \VP{In contrast to the existing studies, here we explore musical content analysis by incorporating linguistic cues of lyrics to give a better overview of how music preferences reflect on moral values.}
% These studies 

% Importantly, these associations were stronger for listeners who generally liked a song because of its lyrics rather than melody. 

% Extroverts tend to listen to songs with positive emotion words, while neurotic individuals favour lyrics expressing more negative emotions. 

% Qiu and colleagues \cite{qiu2019personality} found that 
% conscientiousness was positively correlated with achievement words, extroverts tend to listen to songs with positive emotion words, while neurotic individuals favour lyrics expressing more negative emotions. 

% Applying lyrical content analysis to smartphone music listening data, Sust et al.~\cite{sust2022personality} observed connections between money-themed lyrics and lower levels of openness traits, and between higher levels of conscientiousness and a tendency for love-themed lyrics.

% \subsection{\CS{Lyrics in MIR}}

% \CS{actually not sure we should have a further subsection here.. given that work related to sentiment and emotion analysis is discussed in the subsequent sections. Thoughts?}
% \KK{for me it's already a huge section. I would put these works together with the previous subsection.}

\begin{table}[t]
\centering
\begin{tabular}{@{}lcccc@{}}
\toprule\toprule
 &
   &
  \textbf{Census} &
%   \begin{tabular}[c]{@{}c@{}}\textbf{FB}\\      \textbf{App}\\      $n=$ \\ $63,980$\end{tabular} &
  \begin{tabular}[c]{@{}c@{}}\textbf{MFT}   \\      \textbf{All data}\\     $n=3,920$\end{tabular} &
  \begin{tabular}[c]{@{}c@{}}\textbf{MFT \& $\ge$10}\\       \textbf{Page Likes}\\      $n=1,386$\end{tabular} \\ \midrule
\multirow{2}{*}{\textbf{Gender}} & M             & 48\% &  54\% & 53\% \\
                        & F           & 52\% & 46\% & 47\% \\ \midrule
\multirow{2}{*}{\textbf{Age}}    & $<$25    & 23\%  & 21\% & 29\% \\
                        & $\ge$25 & 77\% & 79\% & 71\% \\\bottomrule\bottomrule
\end{tabular}
\caption{Demographic breakdown of our data according to gender and age. The ``Census'' column reports the national distribution per attribute according to the statistics provided by the official census bureau \cite{istat}.} 
% The ``Dataset'' column reports the percentages of the total number of participants for which we have complete demographic records
% KK{census age does not sum to 100}
\label{tab:initial_demographics}
\end{table}

\section{Data Collection}\label{sec:page_size}

The LikeYouth Facebook-hosted application was initially launched in March 2016, while the data used here were downloaded in September 2019.
% \cite{urbinati2020young}. %Therefore 90\% of the users in this dataset are Italian. 
% Participants after agreeing to share their likes on Facebook pages and basic demographic information were proposed to fill in surveys and quizzes. 
It was deployed mainly in Italy, where approximately 64,000 people entered the platform, from whom 3,920 users (90\% geolocated in Italian territory) filled out the MFT questionnaire correctly.

% Table~\ref{tab:initial_demographics} shows the demographic breakdown of the dataset. 
%Then we filtered out users with less than ten likes for different English-singing artists. 
%We also removed users who didn't reveal at least one of the demographic information (age or gender). Finally, we applied a survey quality check; we removed users that used the same score number for all the questions and users with uncompleted the questionnaires. 
%The process of the data cleaning and preprocessing,left us with 1,386 users. 
% \KK{TODO: predict age missing values from Facebook Likes. You can use the entire dataset.
% Same for gender if this was also inferred.}
%Talk about lyrics and the preprocessing of the text
% mentioned that Language  Mention that we kept only english lyrics.
% Have a diagram with the steps of the preprocessing 
% \noindent \textbf{Data Preprocessing.} 
Of those, 47\% did not provide their age due to the facultative nature of LikeYouth.
Because we wished to include age as a demographic predictor variable, 
we inferred the missing values from all (e.g., not just music artist related) Page Likes of the 3,920 users.
Similar to \cite{kosinski2014manifestations} we created a sparse matrix representation of Page Likes per user and applied sparse singular value decomposition to reduce dimensionality, while binning the age attribute (median $=25$) as ``younger'' ($< 25$) and ``older'' ($\ge 25$)
% \VP{based on the sample median ($<$25 and $\ge$25)}
allowed to approximate the official census distribution \cite{istat}.
We then employed an XGBoost classifier, to predict missing age values \cite{kalimeri2019predicting}, with an estimated 
% accuracy F1-score = 0.89, 
$AUROC=0.79$ and standard deviation $=0.018$. 
Acknowledging that age inference might add bias to our models, we only use age as a predictor in isolated experiments (see Table \ref{tab:independent_features}).
% Further, we conducted the same experiment by keeping only the participants with age information, and we obtained similar prediction results for binding but slightly lower results for individualising.
We also run the same experiments keeping only users who provided their age. Predictions were similar for Binding and slightly lower for Individualising.

% In terms of demographics, we briefly analyse age and gender of participants while focusing more in users' artist page likes and lyrics features extracted from the artists' songs. We also employ the information about the overall number of likes for each artist page to indicate artist popularity. Table \ref{tab:initial_demographics} depicts basic user demographics including the information about overall number of song lyrics extracted from the liked artist. It should be noted that that about 30\% of the participants did not share the age information, hence we applied data imputation for the missing values. This might add noise to the experiments that include age characteristic.

% \begin{figure*}
%  \centerline{
%  \includegraphics[width=2.0\columnwidth]{latex/Data_Structure.png}}
%  \caption{Figure captions should be placed below the figure.}
%  \label{fig:example}
% \end{figure*}

% \begin{figure}
%  \centerline{
%  \includegraphics[width=0.95\columnwidth]{latex/Model_pipeline.png}}
%  \caption{Figure captions should be placed below the figure.}
%  \label{fig:example}
% \end{figure}

%KK: this goes somewhere else but not here. 
%Besides the user demographics and Facebook artist likes we also utilised artist popularity feature which is provided in the LikeYouth dataset.  

To ensure the stability of our regression models, we applied a simple activity threshold. After extensive experimentation we chose to drop users with less than 10 Facebook Page Likes related to music artists (Page category selection), resulting in a reduced final dataset of 1,386 users.
% \VP{We considered the number of artist-liked pages per user (mean = 35.11, sd = 33.95) as an additional feature that can potentially convey additional information about users' music habits. Besides, we employed an already built-in feature from the dataset for representing artist popularity based on the Facebook Page followers.}
% we included in our final dataset only those who had more than 10 Likes on Facebook pages related to music (page category selection), hence remaining with only 1,386 participants. The introduction of this activity threshold, although reducing the sample size significantly is important for the model stability. 
% Table~\ref{tab:initial_demographics} shows the demographic breakdown of the final dataset. 
Table~\ref{tab:initial_demographics} reports the demographic breakdown of our data sample in terms of gender and age,  which follows closely the population distribution of the official Italian census \cite{istat}.
% \KK{\footnote{90\% of the participants we geolocated in the Italian territory.}}.

% \noindent \textbf{Lyrics Information Processing.} 

For the final 1,386 users, we retrieved song lyrics corresponding to their music artist Page Likes using \textit{genius.com}. 
% Querying the Genius API
% , we specifically obtained the top 5 most popular songs per artist alongside the respective lyrics. 
Querying the Genius API, we initially obtained the 10 most popular songs per artist alongside the respective lyrics.
We assume that if a user liked the Page of a specific artist, then that artist's most famous songs (as per Genius) reflect the music preferences of the user.  
% \VP{Among multiple experiments (using 10, 5 and 3 artists' songs), we opted for the 5 most popular songs of a musician. This gave us a decent solution with reasonable computational resources,} which accounts for within-musician variability in lyrical and audio content (see future work discussion) while maintaining an optimal number of musicians and songs for our lyrics data.
We carried out predictive tasks using the $n=$ 10, 5, or 3 most popular songs from an artist and found that $n=$ 5 gave the best compromise in terms of predictions, computational resources, and within-musician variability in lyrical and audio content (see future work discussion) while maintaining an optimal number of musicians and songs for our lyrics data.
Finally, we used the spaCy library \cite{spacy2,neumann-etal-2019-scispacy} to identify songs with English lyrics only, resulting in 3,179 artists and 15,895 songs.

We also considered two additional, more shallow digital trace features that can potentially convey information about user's music habits, namely the number of Page Likes per user (mean $=35.11$, standard deviation $=33.95$) and a built-in feature of artist popularity from LikeYouth, based on the number of Page followers.

% \VP{We use this particular dataset because, to our knowledge, it is the only one that provides participants' moral scores besides their demographics and music preferences. We provide the lyrics data and the source code for lyrics preprocessing and feature modeling \footnote{\url{https://github.com/vjosapreniqi/repo_to_be_added}}. Whereas the moral questionnaires and models cannot be shared due to privacy implications.}

% as an additional feature that can potentially convey information about users' music habits. Besides, we employed an already built-in feature from the dataset for representing artist popularity based on the Facebook Page followers.

% On average, each song contained 120 words.

% It should mentioned that the 80\% of the songs were in English language; the rest in 33 different languages while only the 3\% in Italian language. 

% On average, we had 35 Page Likes per user.

% with lyrics containing 120 words on average. While, the mean of artist Liked pages per user is 35.

We use LikeYouth because, to our best knowledge, it is the only dataset providing MFT scores of individuals alongside a potential proxy of their music preferences (e.g., artist Page Likes).
A limitation of this approach is that the data provided by LikeYouth are static and may thus refer to a snapshot of music interests in time. Streaming platforms  could offer richer information about habitual music listening \cite{anderson2021just,sust2022personality}. Nonetheless, there is substantial evidence that Facebook Page Likes can capture personality needs and personal values \cite{nave2018musical,urbinati2020young,kosinski2014manifestations}.
Another limitation is that LikeYouth user MFT scores and thus our predictive models cannot be made publicly available due to privacy implications \cite{kalimeri2019human}. Instead, we have shared the lyrics data and related source code for lyrical feature modeling in a GitHub repository.\footnote{\url{https://github.com/vjosapreniqi/lyrics-content-features}}

\begin{table}[t]
 \begin{center}
\begin{tabular}{@{}lll@{}} \toprule\toprule
\textbf{Type} & \textbf{Method} & \textbf{Features}  \\ \midrule
Topics & {LDA}                         & \begin{tabular}[c]{@{}l@{}}Death/Fear/Violence, \\  Obscene, Romantic, \\ World/Time/Life\end{tabular} \\ \midrule
Morals & {MoralStrength}               & \begin{tabular}[c]{@{}l@{}}Care, Fairness, Loyalty, \\  Authority, Purity\end{tabular}  \\ \midrule
Sentiment & {VADER}                       & \begin{tabular}[c]{@{}l@{}}Negative,   Positive,\\  Neutral, Compound\end{tabular}  \\ \midrule
Emotions & {NRC} & \begin{tabular}[c]{@{}l@{}} Anger, Disgust, Fear, \\ Sadness, Anticipation, \\ Surprise, Joy, Trust\end{tabular} \\ \bottomrule
\bottomrule
\end{tabular}
\end{center}
 \caption{Summary of lyrical features used in this study.}
 \label{tab:lyrics_tools}
\end{table}

% \KK{compute the topic prevalence}
\begin{table}[t]
 \begin{center}
 \small
\begin{tabular}{@{}llll@{}}
\toprule\toprule
\textbf{Topic} & \textbf{Artist} & \textbf{Song Title} & \textbf{\%} \\ \midrule
\multirow{5}{*}{\begin{tabular}[c]{@{}l@{}}Romantic\\ (0.39)\end{tabular}} & Mike Williams & Give it up & 99 \\
 & Marc Anthony & I need to know & 99 \\
 & NSYNC & I want you back & 99 \\
 & Willie Nelson & Always on my mind & 98 \\
 & Alexia & Because I miss you & 97 \\ \midrule
\multirow{5}{*}{\begin{tabular}[c]{@{}l@{}}Obscene\\ (0.24)\end{tabular}} & Tyga & Rack city & 98 \\
 & Fat Joe & Yellow tape & 96 \\
 & Cardi B & Bartier cardi & 95 \\
  & Chamillionaire & Ridin' & 95 \\
  & 21 Savage & Bank account & 91 \\ \midrule
\multirow{5}{*}{\begin{tabular}[c]{@{}l@{}}World/Time/\\ Life\\ (0.22)\end{tabular}} 
 & Holly Herndon	& Morning sun & 99\\
 & Noisecontrollers & The day & 97 \\
 & Nathan East & Finally home & 96 \\
 & Dave Gahan & Tomorrow & 94 \\
%  & Labyrinth & Miles Away & 93 \\
 & Gabrielle Aplin & Start of time & 90 \\ \midrule

\multirow{5}{*}{\begin{tabular}[c]{@{}l@{}}Death/Fear/\\ Violence\\ (0.15)\end{tabular}}
% & Nitzer Ebb & Join in the chant & 99 \\
 & Hatebreed & Destroy everything & 99\\	
 & Fear Factory & Edgecrusher & 97 \\
 & Eomac & Mandate for murder & 95 \\
 & Destruction & Thrash till death & 92 \\
 & Sabaton & Attack of dead men & 91 \\ 
%  & Immortal & Northern Chaos Gods & 90 \\
 \bottomrule\bottomrule
\end{tabular}
\end{center}
 \caption{LDA topic modelling: overall topic prevalence (in brackets below topic descriptions) and 5 manually selected songs per topic as ranked by descending topic proportion.}
 \label{tab:lyrics_topics}
\end{table}

\section{Lyrics Content Analysis}
 
% \VP{Alternatively I can add a paragraph here to discuss lyrics processing steps and also tell that not all the approaches had the same level of the text processing}

% \VP{In this work, we applied different levels of lyrics preprocessing based on the feature extraction methods. For sentiment analysis, we only applied a general cleanup while keeping the punctuation and capitalization within the text. While for other lyrics feature extraction approaches, we extracted Part Of Speech (POS) lemmas using the spaCy lemmatizer \cite{spacy2}. From the tokenization process, we saw that the lyrics in our dataset contain 274 words and 109 lemmas on average.}

We extracted a set of
textual features related to each song lyrics' overarching narrative (topic modelling), moral valence, sentiment, and emotion. Based on the corresponding feature modeling method, we applied different levels of text preprocessing. Sentiment detection required only a general cleanup while keeping punctuation and capitalization within the text. For the other methods, we extracted Part Of Speech (POS) lemmas using the spaCy lemmatizer \cite{spacy2}. On average, each lyrics contained 273 words and 108 lemmas.

\subsection{Topic Modelling} 
Initially, we aimed to uncover common patterns in the lyrics narratives by applying a topic modelling approach based on Latent Dirichlet Allocation (LDA)~\cite{blei2003latent}.
We used LDA due to its simplicity, high accuracy in topic modelling, and good computational efficiency \cite{lancichinetti2015high}.
The input of the LDA model is a term frequency matrix of the corpus created by the song lyrics. To eliminate very common terms that can lead to irrelevant topics, we ignored words with frequency higher than 90\%.

To derive the optimum number of topics $k$, we optimized the topic coherency ($C_v$ metric~\cite{roder2015exploring}) for models with $k \in [2,16]$ using a step size of 2. 
The number of topics for which coherency was maximised was $k=4$.
For $k>4$, we obtained topics that were either generic or hard to characterise due to the mixture of different words belonging to multiple topics. While for $k=4$, the topics obtained were in line with previous literature \cite{sasaki2014lyricsradar, misael2020temporal}.
Table \ref{tab:lyrics_topics} depicts examples of manually selected songs of 5 artists for each topic, ranked by descending weight in the specific topic. 
% The topics are ranked according to their prevalence.
% \VP{while the songs contain more than 50\% of the words which characterise each topic.} 
% Observing the results, we notice that the obtained topics are in line with previous literature \cite{sasaki2014lyricsradar, misael2020temporal}.

\subsection{Moral Valence}

We assess the moral narratives by employing the MoralStrength lexicon~\cite{araque2020moralstrength}, which holds the state-of-the-art performance in moral text prediction. This expands the Moral Foundation Dictionary by offering three times more moral-annotated lemmas. 
% using the WordNet lexical database \cite{miller1995wordnet}. 
% and giving a set of normative ratings for empirical evaluation of morality that goes beyond the binary nature of MFD \cite{araque2020moralstrength}.
The lexicon provides, along with each lemma, the \textit{moral valence score}, a numeric assessment that indicates both the polarity and the intensity of the lemma in each of the five moral foundations (MFT traits).
Moral valence is expressed on a Likert scale from one to nine, with five considered neutral. When lower than 5, scores reflect notions closer to Harm, Cheating, Betrayal, Subversion, and Degradation, while values higher than 5 indicate Care, Fairness, Loyalty, Authority, and Purity, respectively.

We obtained a moral valence score for each lemma in a song's lyrics and each MFT trait, which is then averaged across lemmas for each song.
Negation correction was not applied, as moral foundation polarities do not directly translate as opposites (e.g., ``not care'' is not the same as ``harm'').
The MoralStrength lexicon has a limited linguistic coverage; as a result, we could not predict moral valence for 16\% of the collected lyrics.
Instead, we assigned them the value 5, the neutral point of the moral valence Likert scale.
This approach pushes the observed mean towards the center of the scale, but captures the variability of the moral values across all the lyrical data.

\subsection{Sentiment and Emotion Analysis}

% We complemented the analysis of the lyrics by estimating their sentiment and emotional content. 
% \CS{Although the two concepts have often been used interchangeably, sentiments are differentiated from emotions by the duration in which they are experienced \cite{munezero2014they}.}
In textual data, emotions, as brief and preconscious phenomena, can be defined via descriptions of appraisal, physiological reaction, expressive display, feeling, or action tendency, while sentiments, as lasting and conscious emotional dispositions, tend to be modelled in terms of text polarity (positive, negative, neutral) \cite{munezero2014they}.
% in text analysis are simplified and defined by polarity \cite{munezero2014they}.

We applied the commonly used VADER (Valence Aware Dictionary and sEntiment Reasoner) model \cite{hutto2014vader} on the lyrical text to obtain information about the sentiment of each song. 
The VADER model is shown to perform well both with long and short text, providing for each song a score for positive, neutral, negative, and compound sentiment (see Table \ref{tab:lyrics_tools}). 
We also estimated the eight basic emotions defined in the Plutchik wheel of emotions \cite{plutchik1982psychoevolutionary} employing the NRC Word–Emotion Association Lexicon \cite{mohammad2013crowdsourcing}. This lexicon was shown to be efficient with unlabeled data~\cite{ccano2017moodylyrics}. 
% The lexicon was applied on the tokenized \textit{Part Of Speech (POS)} lemmas using the spaCy lemmatizer~\cite{spacy2}, while it is shown to be efficient with unlabeled data~\cite{ccano2017moodylyrics}. 
Each song lyrics was annotated with the eight emotions (see Table \ref{tab:lyrics_tools}) by averaging its word emotion association scores.

\section{Experiments and Results}\label{sec:typeset_text}

\begin{figure}[t]
 \centerline{
 \includegraphics[width=1.1\columnwidth]{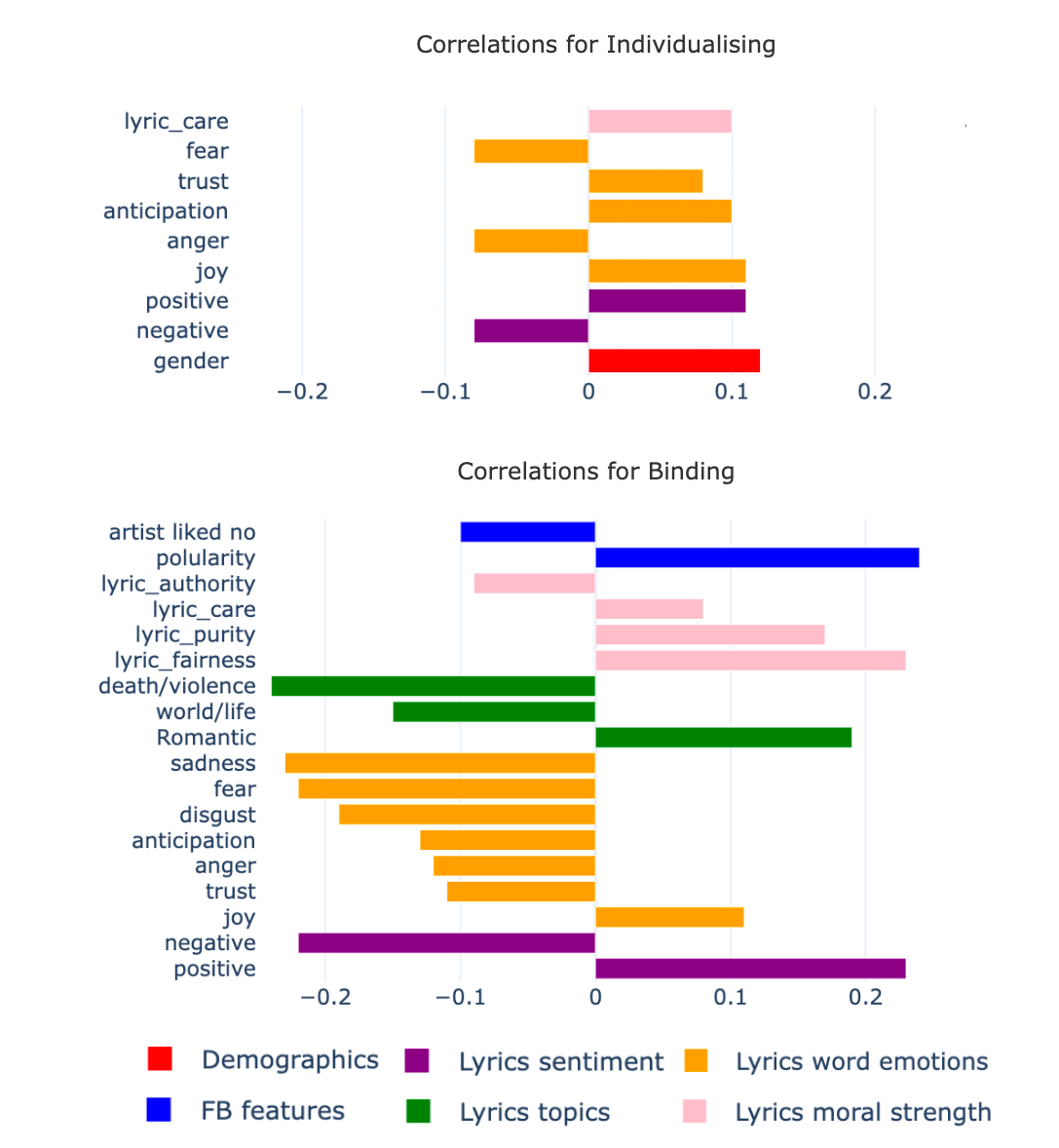}}
 \caption{Spearman's Rank correlations of MFT super foundations with demographics, artist likes and lyrics features.
We report only ones that were significant at $p\leq.01$. 
% Participants high in Individualising foundations liked on average $36,9\pm37,8$ artists, while Binders $33,8\pm35,5$. 
``Artist liked no'' refers to the number of artist Likes per user. 
% (minimum threshold is 10).
% \KK{Also, we need to say what is "artist liked no"}
 }
 \label{fig:super_mft_correlations}
\end{figure}

% \subsection{Correlation analysis}
% \noindent\textbf{Correlation analysis.} 
Initially, we explored the relationship between users' moral values as emerged from the self-reported questionnaires, basic  demographic attributes, and their respective music preferences, as expressed in the linguistic components of the lyrics.
Figure~\ref{fig:super_mft_correlations} depicts the statistically significant correlations ($p\leq.01$) obtained for the two superior foundations, namely Individualising and Binding. 
We observed that people who value more Individualising foundations prefer artists whose songs prevalently talk about anticipation and trust.
On the other side, those concerned more about social order and Binding foundations tend to prefer artists who deal with more romantic topics in their songs instead of existential and social issues.
Overall, participants with strong Binding foundations display a tendency to dislike songs with negative valence and emotions such as sadness, fear, or disgust.
Yet both the Individualising and Binding groups resonate with positive and joyful songs, showing that despite often profound differences in sociopolitical stances, music is a shelter to everyone.

\begin{table}[t]
 \begin{center}
%  \small
 \begin{tabular}{@{}ll@{}}
  \toprule\toprule
    \textbf{ID} & \textbf{Features} \\ \midrule 
    EX1 & Sentiment (VADER) \\ 
    EX2 & Emotions (NRC)  \\ 
    EX3 & Sentiment + Emotions \\ 
    EX4 & Best of \{EX1, EX2, EX3\} + Morals \\ 
    EX5 & Best of \{EX1, EX2, EX3\} + Topics \\ 
    EX6 & Best of \{EX1, EX2, EX3\} + Morals + Topics  \\ 
    EX7 & EX6 + Age + Gender \\
    EX8 & EX7 + Artist Likes + Artist Popularity \\ 
    \bottomrule\bottomrule
 \end{tabular}
\end{center}
 \caption{Summary of performed experiments with corresponding features used as predictors.}
 \label{tab:independent_features}
\end{table}

\begin{table*}[t]
\small
 \begin{center}
\begin{tabular}{@{}lcccccccc@{}}
\toprule\toprule
 \multicolumn{9}{c}{\textbf{Moral Foundations - Regression Models}} \\ 
 \cmidrule{2-9}
 &    EX1  &    EX2 &    EX3  &    EX4  &    EX5 & EX6  &    EX7 &  EX8  \\
\midrule
C          &  .08 [.08, .09] &  .10 [.10, .11] &  .10 [.10, .11] &  .11 [.11, .12] &  .10 [.10, .11] & \textbf{ .12 [.11, .12]} &  \textbf{.12 [.12, .13]} &  .11 [.11, .12] \\
F      &  .04 [.04, .05] &  .06 [.05, .06] &  .05 [.04, .05] &  .06 [.05, .06] &  .06 [.05, .06] &  .05 [.05, .05] &  \textbf{.08 [.07, .08]} &  .05 [.05, .05] \\
L       &  .12 [.12, .13] &  .16 [.16, .17] &  .18 [.17, .18] &  \textbf{.20 [.20, .21]} &  .19 [.18, .19] &  .19 [.19, .20] &  \textbf{.20 [.20, .21] }& \textbf{ .20 [.20, .21]} \\
A     &  .19 [.19, .19] &  .21 [.21, .22] &  .23 [.23, .24] &  .26 [.26, .26] &  .24 [.24, .24] &  .25 [.25, .26] &  .26 [.26, .26] & \textbf{ .27 [.26, .27]} \\
P        &  .19 [.18, .19] &  .20 [.20, .21] &  .24 [.23, .24] &  .25 [.25, .26] &  .23 [.22, .23] &  .25 [.24, .25] &  .24 [.24, .25] & \textbf{ .26 [.26, .26]} \\
\midrule
I &  .08 [.07, .08] &  .10 [.10, .11] &  .10 [.10, .11] &  .10 [.09, .10] &  .09 [.09, .10] &  .10 [.10, .11] & \textbf{ .11 [.10, .11]} &  .10 [.10, .11] \\
B &  .20 [.19, .20] &  .24 [.23, .24] &  .26 [.26, .27] &  .28 [.28, .29] &  .26 [.26, .27] &  .28 [.27, .28] &  .27 [.27, .28] &  \textbf{.30 [.30, .31]} \\
\bottomrule\bottomrule
\end{tabular}
\end{center}
 \caption{Moral foundations regression with Random Forest using different feature combinations (see Table \ref{tab:independent_features}): Pearson's correlation [95\% confidence intervals] between predicted and the actual values averaged across 5-fold cross-validation. C: Care; F: Fairness; L: Loyalty; A: Authority; P: Purity; I: Individualising; B: Binding.}
 \label{tab:mft_regression_results}
\end{table*}

\begin{figure*}[!ht]
 \includegraphics[width=\textwidth]{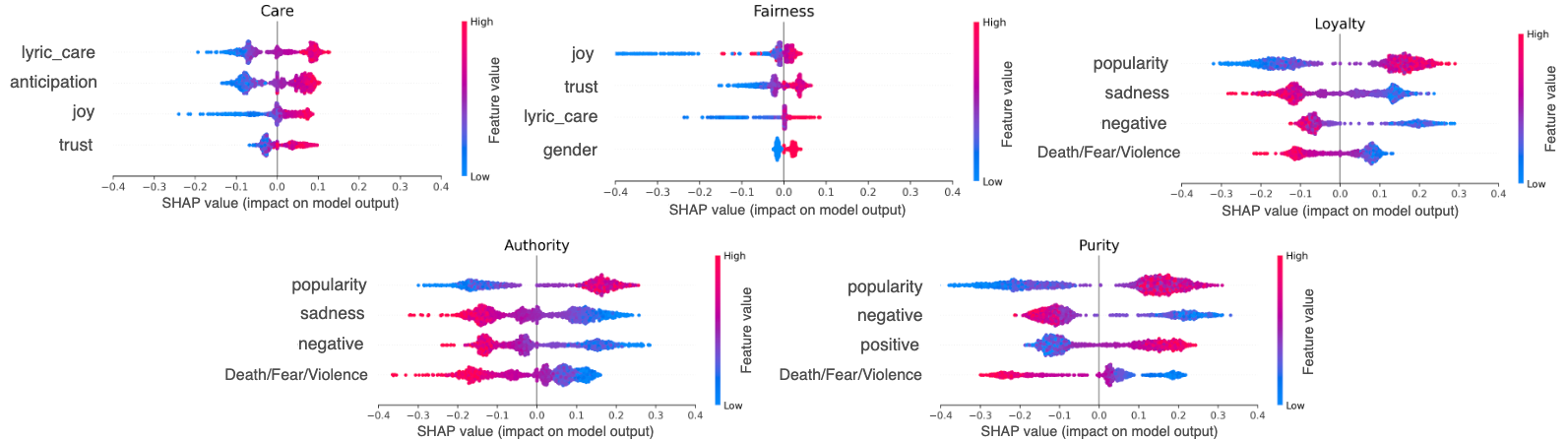}
 \caption{Top 4 individual feature contributions (via SHAP values) for the five basic moral foundations from experiment EX8 (see Table \ref{tab:independent_features}). The higher the SHAP value, the more the feature contributes to the prediction model. 
%  Gender is coded as '0' for males and '1' for females.
 %\KK{gender needs to be explained. what is gender 0.05? either in the figure or in the caption we need to write how it's coded.  }
 %\VP{gender value represents Males and Females in a binary fashion (0,1) in the graph we see that 0 (males) predict lower values of care and 1(females represented with red/high val) predicts higher values of care. Artist liked number is represent the sum of likes per user for different artist pages (the minimum is 10 likes) }
}
 \label{fig:5_MFT_SHAP_values}
\end{figure*}

% \subsection{Regression Modelling}
% \noindent\textbf{Regression Modelling.} 
Next, we proposed a series of experimental designs to infer moral values of the participants from their music preferences and the respective linguistic content. Table~\ref{tab:independent_features} summarises the performed experiments. 
We employed four algorithms, namely Support Vector Regressor, Random Forest, XGBoost, and ElasticNet, to predict moral values using a multivariate regression approach over a 5-fold cross-validation setting. For each participant, the features were aggregated and normalised. Here, we report only the results from the Random Forest since it slightly outperformed the rest. 
We used the Pearson's correlation coefficients between the predicted and actual moral values scores to measure the model's goodness fit.  This metric was commonly used in papers that predicted personality based on users' music preferences and listening behaviours \cite{nave2018musical, anderson2020algorithmic}.

To comprehend the general behaviour of our models and evaluate the importance of each feature, we estimated the SHAP values. SHAP (SHapley Additive exPlanations) is a game theory approach designed to illustrate the features' contribution to the final output of any machine learning model \cite{lundberg2017unified}. 
% SHAP values present both global and local interpretability, meaning that we can assess both how much each predictor and each observation, respectively, contribute to the performance of the regressor. 

% Moving further to the inference of people's moral values from their music preferences, 
% we followed 
Following the incremental experimental design reported in Table \ref{tab:independent_features}, 
we trained one model per each moral foundation and presented the best results obtained by each feature in Table \ref{tab:mft_regression_results}.
In line with recent literature~\cite{preniqi2021modelling} that shows higher prediction accuracy for Binding rather than Individualising foundations, we noticed a similar behaviour also when inferring from linguistic features of song lyrics.

When adding demographics and artist Facebook information (EX8), the results slightly improved for both super foundations, implying that the more information we have about users' demographics and music preferences, the more precise our models become.
Despite that, the model trained on just emotions, sentiment and moral information (EX4) achieved almost as good results as those who are aware of the demographics and the general artist information (EX7 and EX8).
This highlights the importance of music preferences in portraying our goals and decisions whose motivations go far beyond basic demographic knowledge.  

\begin{figure*}[!ht]
\centering
\hspace{0.62cm}
\includegraphics[width=0.96\textwidth]{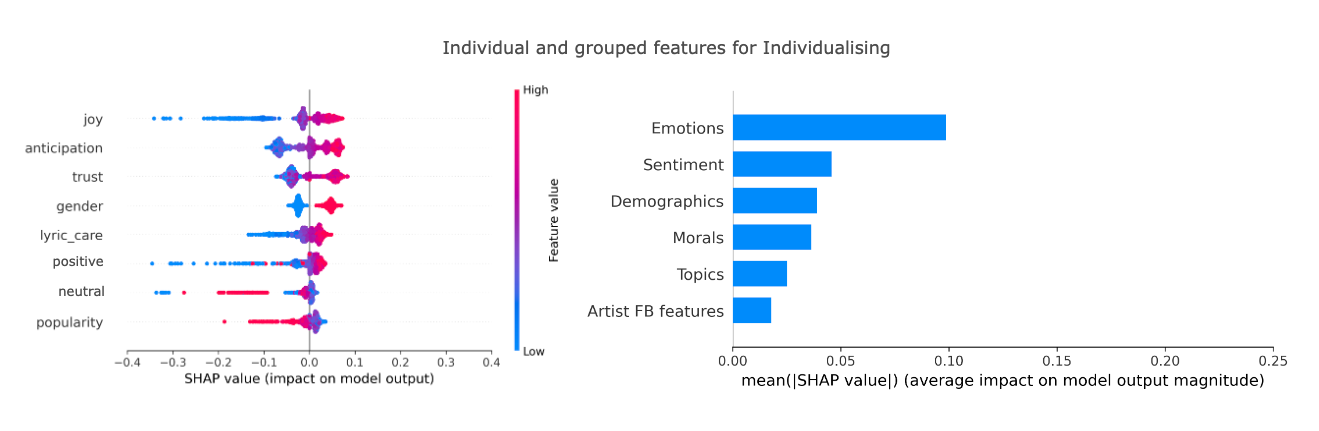}
\includegraphics[width=0.96\textwidth]{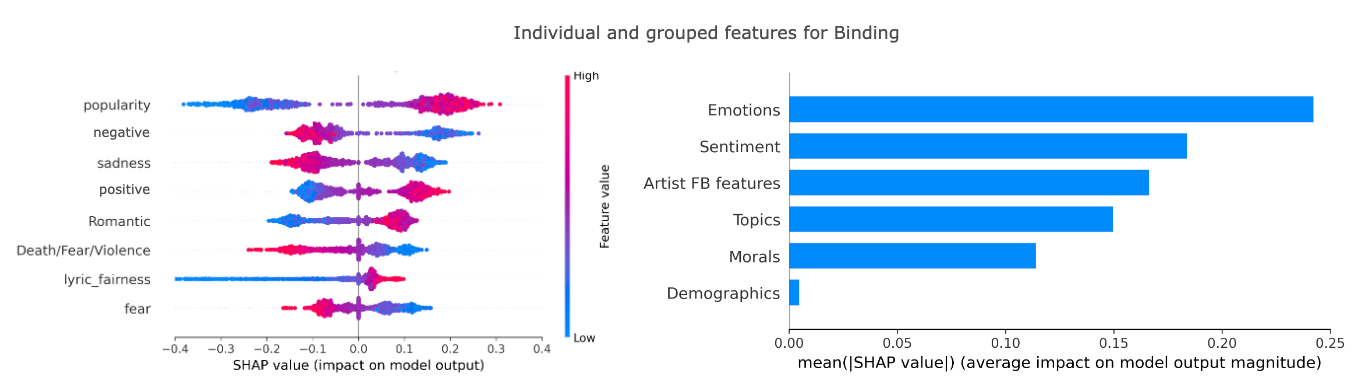}
\caption{Individual (top 8) and grouped feature contributions (via SHAP values) for the two superior moral foundations from experiment EX8 (see Table \ref{tab:independent_features}). The higher the SHAP value, the more the feature contributes to the prediction model.}
 \label{fig:superior_SHAP_values}
\end{figure*}

Figure \ref{fig:5_MFT_SHAP_values} depicts the most important individual (only top 4 due to page restrictions) features for predicting each of the five moral foundations. While Figure \ref{fig:superior_SHAP_values} illustrates the impact individual (top 8) and grouped features in inferring the two superior foundations when considering all predictor variables (EX8). 
% \KK{Due to page restrictions, we report only the four most important predictors for the basic foundations(Figure \ref{fig:5_MFT_SHAP_values}), while for the superior foundations we report the top eight (Figure \ref{fig:superior_SHAP_values})}. 
In line with observed correlations, feature importance representations for regression models show that lyrics linked to objective and subtle emotions (e.g., joy, trust, and anticipation) effectively predict Care and Fairness. Whereas more intense and opposite polarities of sentiment and emotions (e.g., fear, sadness, lyrics positive and negative valence) account for better predictions of Loyalty, Authority and Purity.
We noticed that those who value more the Binding foundations 
% (Loyalty, Purity, and Authority) 
appear to be sensitive to the popularity of the song, which reflects their worldview of prioritising group-focus over self-focus.
% the needs of their social group, sacrificing their self-interests.}

% \begin{figure*}[!ht]
%  \centerline{
%  \includegraphics[width=2.1\columnwidth]{latex/figs/Individualising_grouped_features.png}}
%  \caption{Individual and grouped feature contributions (via SHAP values) for Individualising foundation. The higher the SHAP value, the more the feature contributes to the moral prediction}
%  \label{fig:Individualising_SHAP_values}
% \end{figure*}

% \begin{figure*}[!ht]
%  \centerline{
%  \includegraphics[width=2.1\columnwidth]{latex/figs/Binding_grouped_features.png}}
%  \caption{Individual and grouped feature contributions (via SHAP values) for Binding foundations. The higher the SHAP value, the more the feature contributes to the moral prediction}
%  \label{fig:Binding_SHAP_values}
% \end{figure*}

\section{Conclusion}

This paper discussed the link between lyrical information and moral values. We presented a wide range of lyrics processing techniques and features for measuring the power of linguistic aspects in predicting complex psychological traits such as moral values. Besides, we explored and compared the impact of user demographics and shallow digital traces 
in inferring moral foundations against the song lyrics components. 

We noticed that Binders express their views throughout their music preference and lyrical styles. In contrast, Individualising views are more complex to be captured solely by people's music lyrics preferences. 
% This implies that people with higher scores towards values of Binding, more instinctively express their moral values through the music they listen to. 
Thus, using the proposed framework, it was easier to infer moral values of Binding ($.20 \leq r \leq .30$)  between predicted and target values than Individualising foundations ($.08 \leq r \leq .11$). 

%This paragraph most likely should be moved to section 5.2

We demonstrated that lyrics features extracted from the naturally emerging music preferences in social media, to some extent, allow for constructing reliable inferences of moral values. Considering the expanded presence of online music streaming services our findings may have direct implications for music recommendation and personalisation algorithms \cite{nave2018musical, ogden2011music, porcaro2021diversity}. Since moral values are a key element of the decision making process in several societal issues \cite{haidt2004intuitive, kalimeri2019predicting} and highly linked to political leanings \cite{lakoff2010moral}, our research implications can help future studies to tackle aspects of why and how music is or can be used for mass stimulation and persuasion in social and political campaigns, raising awareness on what our digital music behaviours can reveal.

% \VP{This can contribute in raising awareness about how much information can be encoded in our everyday behaviours, such as music listening preferences.}

% However, the results call for further investigation as there is space for improvement.
% These outcomes emerged as a result of an exploitation of lyrics sentiment, emotions, moral valence, and topics from the most popular songs of the artists liked on Facebook which allowed us to conduct a comparative study on the predictive power of each modality and their combination. Further, we saw that people tend to match their moral values with the moral valence inferred by the lyrics of the songs (except for the authority foundation). Such insights are crucial and can help fine-tune future music recommendation algorithms.

In future work, we intend to combine audio and lyrical content analysis together in a multimodal framework to further expand our understanding of music and moral affiliations, especially for Individualising foundations that remain hard to predict. 
Recent work highlights that preferences for both lyrics and audio features are important in predicting, often distinctly, personality traits \cite{sust2022personality}.
We will also use additional data from LikeYouth to investigate if moral foundations can explain variance in music preferences that cannot be accounted for by personality traits and personal values (cf.~\cite{messick2020role}).
We ultimately aim to integrate our findings into novel psychologically aware music recommender systems, but also beyond the music domain to other media.

\section{Acknowledgements}

This work was supported by the QMUL Centre for Doctoral Training in Data-informed Audience-centric Media Engineering (2021–2025) as part of a PhD studentship awarded to VP.
KK acknowledges support from the Lagrange Project of the Institute for Scientific Interchange Foundation (ISI Foundation) funded by Fondazione Cassa di Risparmio di Torino (Fondazione CRT). We would like to thank the three anonymous reviewers and the meta-reviewer for their thoughtful comments.

\FloatBarrier
% For bibtex users:
% \bibliography{ismir}

\begin{thebibliography}{ismir}
% Example bibtex file for ISMIR Template
% \bibitem{Author:17}
% E.~Author and B.~Authour, ``The title of the conference paper,'' in {\em Proc.
% of the Int. Society for Music Information Retrieval Conf.}, (Suzhou, China),
% pp.~111--117, 2017.
%
% \bibitem{Someone:10}
% A.~Someone, B.~Someone, and C.~Someone, ``The title of the journal paper,''
%  {\em Journal of New Music Research}, vol.~A, pp.~111--222, September 2010.
%
% \bibitem{Person:20}
% O.~Person, {\em Title of the Book}.
% \newblock Montr\'{e}al, Canada: McGill-Queen's University Press, 2021.
%
% \bibitem{Person:09}
% F.~Person and S.~Person, ``Title of a chapter this book,'' in {\em A Book
% Containing Delightful Chapters} (A.~G. Editor, ed.), pp.~58--102, Tokyo,
% Japan: The Publisher, 2009.
%
%
% \providecommand{\url}[1]{#1}
\csname url@samestyle\endcsname
\providecommand{\newblock}{\relax}
\providecommand{\bibinfo}[2]{#2}
\providecommand{\BIBentrySTDinterwordspacing}{\spaceskip=0pt\relax}
\providecommand{\BIBentryALTinterwordstretchfactor}{4}
\providecommand{\BIBentryALTinterwordspacing}{\spaceskip=\fontdimen2\font plus
\BIBentryALTinterwordstretchfactor\fontdimen3\font minus
  \fontdimen4\font\relax}
\providecommand{\BIBforeignlanguage}[2]{{%
\expandafter\ifx\csname l@#1\endcsname\relax
\typeout{** WARNING: IEEEtran.bst: No hyphenation pattern has been}%
\typeout{** loaded for the language `#1'. Using the pattern for}%
\typeout{** the default language instead.}%
\else
\language=\csname l@#1\endcsname
\fi
#2}}
\providecommand{\BIBdecl}{\relax}
\BIBdecl

\bibitem{porcaro2021diversity}
L.~Porcaro, C.~Castillo, and E.~G{\'o}mez~Guti{\'e}rrez, ``Diversity by design
  in music recommender systems,'' \emph{Transactions of the International
  Society for Music Information Retrieval. 2021; 4 (1).}, 2021.

\bibitem{laplante2014improving}
A.~Laplante, ``Improving music recommender systems: What can we learn from
  research on music tastes?'' in \emph{ISMIR}, 2014, pp. 451--456.

\bibitem{rentfrow2003re}
P.~J. Rentfrow and S.~D. Gosling, ``The do re mi's of everyday life: the
  structure and personality correlates of music preferences.'' \emph{Journal of
  personality and social psychology}, vol.~84, no.~6, p. 1236, 2003.

\bibitem{greenberg2016song}
D.~M. Greenberg, M.~Kosinski, D.~J. Stillwell, B.~L. Monteiro, D.~J. Levitin,
  and P.~J. Rentfrow, ``The song is you: Preferences for musical attribute
  dimensions reflect personality,'' \emph{Social Psychological and Personality
  Science}, vol.~7, no.~6, pp. 597--605, 2016.

\bibitem{nave2018musical}
G.~Nave, J.~Minxha, D.~M. Greenberg, M.~Kosinski, D.~Stillwell, and
  J.~Rentfrow, ``Musical preferences predict personality: evidence from active
  listening and facebook likes,'' \emph{Psychological science}, vol.~29, no.~7,
  pp. 1145--1158, 2018.

\bibitem{devenport2019predicting}
S.~P. Devenport and A.~C. North, ``Predicting musical taste: Relationships with
  personality aspects and political orientation,'' \emph{Psychology of Music},
  vol.~47, no.~6, pp. 834--847, 2019.

\bibitem{anderson2021just}
I.~Anderson, S.~Gil, C.~Gibson, S.~Wolf, W.~Shapiro, O.~Semerci, and D.~M.
  Greenberg, ``“just the way you are”: Linking music listening on spotify
  and personality,'' \emph{Social Psychological and Personality Science},
  vol.~12, no.~4, pp. 561--572, 2021.

\bibitem{gardikiotis2012rock}
A.~Gardikiotis and A.~Baltzis, ``‘rock music for myself and justice to the
  world!’: Musical identity, values, and music preferences,''
  \emph{Psychology of Music}, vol.~40, no.~2, pp. 143--163, 2012.

\bibitem{swami2013metalheads}
V.~Swami, F.~Malpass, D.~Havard, K.~Benford, A.~Costescu, A.~Sofitiki, and
  D.~Taylor, ``Metalheads: The influence of personality and individual
  differences on preference for heavy metal.'' \emph{Psychology of Aesthetics,
  Creativity, and the Arts}, vol.~7, no.~4, p. 377, 2013.

\bibitem{manolios2019influence}
S.~Manolios, A.~Hanjalic, and C.~C. Liem, ``The influence of personal values on
  music taste: Towards value-based music recommendations,'' in
  \emph{Proceedings of the 13th ACM Conference on Recommender Systems}, 2019,
  pp. 501--505.

\bibitem{hu2011enhancing}
R.~Hu and P.~Pu, ``Enhancing collaborative filtering systems with personality
  information,'' in \emph{Proceedings of the fifth ACM conference on
  Recommender systems}, 2011, pp. 197--204.

\bibitem{jin2020effects}
Y.~Jin, N.~Tintarev, N.~N. Htun, and K.~Verbert, ``Effects of personal
  characteristics in control-oriented user interfaces for music recommender
  systems,'' \emph{User Modeling and User-Adapted Interaction}, vol.~30, no.~2,
  pp. 199--249, 2020.

\bibitem{lu2018diversity}
F.~Lu and N.~Tintarev, ``A diversity adjusting strategy with personality for
  music recommendation.'' in \emph{IntRS@ RecSys}, 2018, pp. 7--14.

\bibitem{loersch2013unraveling}
C.~Loersch and N.~L. Arbuckle, ``Unraveling the mystery of music: Music as an
  evolved group process.'' \emph{Journal of Personality and Social Psychology},
  vol. 105, no.~5, p. 777, 2013.

\bibitem{savage2021music}
P.~E. Savage, P.~Loui, B.~Tarr, A.~Schachner, L.~Glowacki, S.~Mithen, and W.~T.
  Fitch, ``Music as a coevolved system for social bonding,'' \emph{Behavioral
  and Brain Sciences}, vol.~44, 2021.

\bibitem{demetriou2018vocals}
A.~M. Demetriou, A.~Jansson, A.~Kumar, and R.~M. Bittner, ``Vocals in music
  matter: the relevance of vocals in the minds of listeners.'' in \emph{ISMIR},
  2018, pp. 514--520.

\bibitem{ali2006songs}
S.~O. Ali and Z.~F. Peynircio{\u{g}}lu, ``Songs and emotions: are lyrics and
  melodies equal partners?'' \emph{Psychology of music}, vol.~34, no.~4, pp.
  511--534, 2006.

\bibitem{lee2021a}
J.~H. Lee, A.~Bhattacharya, R.~Antony, N.~K. Santero, and A.~Le, ``Finding
  home: Understanding how music supports listener mental health through a case
  study of bts,'' in \emph{ISMIR}, 2021, pp. 358--365.

\bibitem{qiu2019personality}
L.~Qiu, J.~Chen, J.~Ramsay, and J.~Lu, ``Personality predicts words in favorite
  songs,'' \emph{Journal of Research in Personality}, vol.~78, pp. 25--35,
  2019.

\bibitem{urbinati2020young}
A.~Urbinati, K.~Kalimeri, A.~Bonanomi, A.~Rosina, C.~Cattuto, and D.~Paolotti,
  ``Young adult unemployment through the lens of social media: Italy as a case
  study,'' in \emph{International Conference on Social Informatics}.\hskip 1em
  plus 0.5em minus 0.4em\relax Springer, 2020, pp. 380--396.

\bibitem{hutto2014vader}
C.~Hutto and E.~Gilbert, ``Vader: A parsimonious rule-based model for sentiment
  analysis of social media text,'' in \emph{Proceedings of the international
  AAAI conference on web and social media}, vol.~8, no.~1, 2014, pp. 216--225.

\bibitem{mohammad2013crowdsourcing}
S.~M. Mohammad and P.~D. Turney, ``Crowdsourcing a word--emotion association
  lexicon,'' \emph{Computational intelligence}, vol.~29, no.~3, pp. 436--465,
  2013.

\bibitem{araque2020moralstrength}
O.~Araque, L.~Gatti, and K.~Kalimeri, ``Moralstrength: Exploiting a moral
  lexicon and embedding similarity for moral foundations prediction,''
  \emph{Knowledge-based systems}, vol. 191, p. 105184, 2020.

\bibitem{blei2003latent}
D.~M. Blei, A.~Y. Ng, and M.~I. Jordan, ``Latent dirichlet allocation,''
  \emph{Journal of machine Learning research}, vol.~3, no. Jan, pp. 993--1022,
  2003.

\bibitem{preniqi2021modelling}
V.~Preniqi, K.~Kalimeri, and C.~Saitis, ``Modelling moral traits with music
  listening preferences and demographics,'' \emph{arXiv preprint
  arXiv:2107.00349}, 2021.

\bibitem{messick2020role}
K.~J. Messick and B.~E. Aranda, ``The role of moral reasoning \& personality in
  explaining lyrical preferences,'' \emph{PLoS one}, vol.~15, no.~1, p.
  e0228057, 2020.

\bibitem{mejova2019effect}
Y.~Mejova and K.~Kalimeri, ``Effect of values and technology use on exercise:
  implications for personalized behavior change interventions,'' in
  \emph{Proceedings of the 27th ACM Conference on User Modeling, Adaptation and
  Personalization}, 2019, pp. 36--45.

\bibitem{graham2013moral}
J.~Graham, J.~Haidt, S.~Koleva, M.~Motyl, R.~Iyer, S.~P. Wojcik, and P.~H.
  Ditto, ``Moral foundations theory: The pragmatic validity of moral
  pluralism,'' in \emph{Advances in experimental social psychology}.\hskip 1em
  plus 0.5em minus 0.4em\relax Elsevier, 2013, vol.~47, pp. 55--130.

\bibitem{mcadams2006new}
D.~P. McAdams and J.~L. Pals, ``A new big five: fundamental principles for an
  integrative science of personality.'' \emph{American psychologist}, vol.~61,
  no.~3, p. 204, 2006.

\bibitem{miles2015morality}
A.~Miles and S.~Vaisey, ``Morality and politics: Comparing alternate
  theories,'' \emph{Social Science Research}, vol.~53, pp. 252--269, 2015.

\bibitem{haidt2007morality}
J.~Haidt and J.~Graham, ``When morality opposes justice: Conservatives have
  moral intuitions that liberals may not recognize,'' \emph{Social Justice
  Research}, vol.~20, no.~1, pp. 98--116, 2007.

\bibitem{wolsko2016red}
C.~Wolsko, H.~Ariceaga, and J.~Seiden, ``Red, white, and blue enough to be
  green: Effects of moral framing on climate change attitudes and conservation
  behaviors,'' \emph{Journal of Experimental Social Psychology}, vol.~65, pp.
  7--19, 2016.

\bibitem{amin2017association}
A.~B. Amin, R.~A. Bednarczyk, C.~E. Ray, K.~J. Melchiori, J.~Graham, J.~R.
  Huntsinger, and S.~B. Omer, ``Association of moral values with vaccine
  hesitancy,'' \emph{Nature Human Behaviour}, vol.~1, no.~12, pp. 873--880,
  2017.

\bibitem{kalimeri2019human}
K.~Kalimeri, M.~G.~Beir{\'o}, A.~Urbinati, A.~Bonanomi, A.~Rosina, and
  C.~Cattuto, ``Human values and attitudes towards vaccination in social
  media,'' in \emph{Companion Proceedings of The 2019 World Wide Web
  Conference}, 2019, pp. 248--254.

\bibitem{kalimeri2019predicting}
K.~Kalimeri, M.~G. Beir{\'o}, M.~Delfino, R.~Raleigh, and C.~Cattuto,
  ``Predicting demographics, moral foundations, and human values from digital
  behaviours,'' \emph{Comput. Hum. Behav.}, vol.~92, pp. 428--445, 2019.

\bibitem{kim2013moral}
E.~Kim, R.~Iyer, J.~Graham, Y.-H. Chang, and R.~Maheswaran, ``Moral values from
  simple game play,'' in \emph{International Conference on Social Computing,
  Behavioral-Cultural Modeling, and Prediction}.\hskip 1em plus 0.5em minus
  0.4em\relax Springer, 2013, pp. 56--64.

\bibitem{lin2018acquiring}
Y.~Lin, J.~Hoover, G.~Portillo-Wightman, C.~Park, M.~Dehghani, and H.~Ji,
  ``Acquiring background knowledge to improve moral value prediction,'' in
  \emph{2018 ieee/acm international conference on advances in social networks
  analysis and mining (asonam)}.\hskip 1em plus 0.5em minus 0.4em\relax IEEE,
  2018, pp. 552--559.

\bibitem{graham2009liberals}
J.~Graham, J.~Haidt, and B.~A. Nosek, ``Liberals and conservatives rely on
  different sets of moral foundations.'' \emph{Journal of personality and
  social psychology}, vol.~96, no.~5, p. 1029, 2009.

\bibitem{greenberg2020self}
D.~M. Greenberg, S.~C. Matz, H.~A. Schwartz, and K.~R. Fricke, ``The
  self-congruity effect of music.'' \emph{Journal of Personality and Social
  Psychology}, 2020.

\bibitem{lightman2007using}
E.~J. Lightman, P.~M. McCarthy, D.~F. Dufty, and D.~S. McNamara, ``Using
  computational text analysis tools to compare the lyrics of suicidal and
  non-suicidal songwriters,'' in \emph{Proceedings of the Annual Meeting of the
  Cognitive Science Society}, vol.~29, no.~29, 2007.

\bibitem{shriram2021much}
J.~Shriram, S.~Paruchuri, and V.~Alluri, ``How much do lyrics matter? analysing
  lyrical simplicity preferences for individuals at risk of depression,''
  \emph{arXiv preprint arXiv:2109.07227}, 2021.

\bibitem{sust2022personality}
L.~Sust, G.~Kudchadker, R.~Schoedel, T.~Schuwerk, M.~B{\"u}hner, and C.~Stachl,
  ``Personality computing with naturalistic music listening data,'' 
  \emph{PsyArXiv Preprints}, 2022.

\bibitem{dewall2011tuning}
C.~N. DeWall, R.~S. Pond~Jr, W.~K. Campbell, and J.~M. Twenge, ``Tuning in to
  psychological change: Linguistic markers of psychological traits and emotions
  over time in popular us song lyrics.'' \emph{Psychology of Aesthetics,
  Creativity, and the Arts}, vol.~5, no.~3, p. 200, 2011.

\bibitem{mcauslan2018billboard}
P.~McAuslan and M.~Waung, ``Billboard hot 100 songs: Self-promoting over the
  past 20 years.'' \emph{Psychology of Popular Media Culture}, vol.~7, no.~2,
  p. 171, 2018.

\bibitem{istat}
I.Stat, ``Italian statistics,'' \url{http://dati.istat.it/}, accessed: 2019-09.

\bibitem{kosinski2014manifestations}
M.~Kosinski, Y.~Bachrach, P.~Kohli, D.~Stillwell, and T.~Graepel,
  ``Manifestations of user personality in website choice and behaviour on
  online social networks,'' \emph{Machine learning}, vol.~95, no.~3, pp.
  357--380, 2014.

\bibitem{spacy2}
M.~Honnibal and I.~Montani, ``{spaCy 2}: Natural language understanding with
  {B}loom embeddings, convolutional neural networks and incremental parsing,''
  2017, to appear.

\bibitem{neumann-etal-2019-scispacy}
\BIBentryALTinterwordspacing
M.~Neumann, D.~King, I.~Beltagy, and W.~Ammar, ``{S}cispa{C}y: Fast and robust
  models for biomedical natural language processing,'' in \emph{Proceedings of
  the 18th BioNLP Workshop and Shared Task}.\hskip 1em plus 0.5em minus
  0.4em\relax Florence, Italy: Association for Computational Linguistics, Aug.
  2019, pp. 319--327. [Online]. Available:
  \url{https://aclanthology.org/W19-5034}
\BIBentrySTDinterwordspacing

\bibitem{lancichinetti2015high}
A.~Lancichinetti, M.~I. Sirer, J.~X. Wang, D.~Acuna, K.~K{\"o}rding, and
  L.~A.~N. Amaral, ``High-reproducibility and high-accuracy method for
  automated topic classification,'' \emph{Physical Review X}, vol.~5, no.~1, p.
  011007, 2015.

\bibitem{roder2015exploring}
M.~R{\"o}der, A.~Both, and A.~Hinneburg, ``Exploring the space of topic
  coherence measures,'' in \emph{WSDM}, 2015, pp. 399--408.

\bibitem{sasaki2014lyricsradar}
S.~Sasaki, K.~Yoshii, T.~Nakano, M.~Goto, and S.~Morishima, ``Lyricsradar: A
  lyrics retrieval system based on latent topics of lyrics.'' in \emph{Ismir},
  2014, pp. 585--590.

\bibitem{misael2020temporal}
L.~Misael, C.~Forster, E.~Fontelles, V.~Sampaio, and M.~Fran{\c{c}}a,
  ``Temporal analysis and visualisation of music,'' in \emph{Anais do XVII
  Encontro Nacional de Intelig{\^e}ncia Artificial e Computacional}.\hskip 1em
  plus 0.5em minus 0.4em\relax SBC, 2020, pp. 507--518.

\bibitem{munezero2014they}
M.~Munezero, C.~S. Montero, E.~Sutinen, and J.~Pajunen, ``Are they different?
  affect, feeling, emotion, sentiment, and opinion detection in text,''
  \emph{IEEE transactions on affective computing}, vol.~5, no.~2, pp. 101--111,
  2014.

\bibitem{plutchik1982psychoevolutionary}
R.~Plutchik, ``A psychoevolutionary theory of emotions,'' 1982.

\bibitem{ccano2017moodylyrics}
E.~{\c{C}}ano and M.~Morisio, ``Moodylyrics: A sentiment annotated lyrics
  dataset,'' in \emph{Proceedings of the 2017 International Conference on
  Intelligent Systems, Metaheuristics \& Swarm Intelligence}, 2017, pp.
  118--124.

\bibitem{anderson2020algorithmic}
A.~Anderson, L.~Maystre, I.~Anderson, R.~Mehrotra, and M.~Lalmas, ``Algorithmic
  effects on the diversity of consumption on spotify,'' in \emph{Proceedings of
  The Web Conference 2020}, 2020, pp. 2155--2165.

\bibitem{lundberg2017unified}
S.~M. Lundberg and S.-I. Lee, ``A unified approach to interpreting model
  predictions,'' \emph{Advances in neural information processing systems},
  vol.~30, 2017.

\bibitem{ogden2011music}
J.~R. Ogden, D.~T. Ogden, and K.~Long, ``Music marketing: A history and
  landscape,'' \emph{Journal of Retailing and Consumer Services}, vol.~18,
  no.~2, pp. 120--125, 2011.

\bibitem{haidt2004intuitive}
J.~Haidt and C.~Joseph, ``Intuitive ethics: How innately prepared intuitions
  generate culturally variable virtues,'' \emph{Daedalus}, vol. 133, no.~4, pp.
  55--66, 2004.

\bibitem{lakoff2010moral}
G.~Lakoff, \emph{Moral politics: How liberals and conservatives think}.\hskip
  1em plus 0.5em minus 0.4em\relax University of Chicago Press, 2010.
  
\end{thebibliography}
% \bibliographystyle{plain}

% For non bibtex users:

\end{document}